\documentclass[12pt]{iopart}

\usepackage{iopams}  
\begin{document}

\title{Space-time structure may be topological and not geometrical}

\author{Gabriele Carcassi, Christine A. Aidala}

\address{Physics Department, University of Michigan, 450 Church Street\\
	Ann Arbor, MI 48109-1040,
	United States}
\ead{carcassi@umich.edu}
\vspace{10pt}
\begin{indented}
\item[]February 2020
\end{indented}

\begin{abstract}
In a previous effort we have created a framework that explains why topological structures naturally arise within a scientific theory; namely, they capture the requirements of experimental verification. This is particularly interesting because topological structures are at the foundation of geometrical structures, which play a fundamental role within modern mathematical physics. In this paper we will show a set of necessary and sufficient conditions under which those topological structures lead to real quantities and manifolds, which are a typical requirement for geometry. These conditions will provide a physically meaningful procedure that is the physical counter-part of the use of Dedekind cuts in mathematics. We then show that those conditions are unlikely to be met at Planck scale, leading to a breakdown of the concept of ordering. This would indicate that the mathematical structures required to describe space-time at that scale, while still topological, may not be geometrical.
\end{abstract}

%

%
\submitto{\PS}
%
%
%

\section{Introduction}
In our ongoing project, Assumptions of Physics, the idea is to find a minimal set of assumptions from which the different basic theories can be derived. The idea is that, by doing this as formally as possible, we are forced to specify all our implicit starting points, thus clarifying what could possibly be done differently. While other approaches start with a similar goal, see for example Refs.~\cite{PhysRevA.84.012311,QLogicReview,Hardy:2001jk,ludwig_hein_2013}, there are a number of key differences. One difference is that in our work both classical and quantum cases are derived on equal footing, identifying the key point of divergence between the two theories.\cite{Carc1} Another difference is that our approach aims to start with primitives that are necessary to do physics. Our basic building block is the notion of a verifiable statement: an assertion for which an experimental test is available that would confirm, in finite time, that the assertion is true. As one can imagine, without such a notion, we would have no experimental verification, which means no possibility of doing science. In general, the goal is to start with a tight connection between the physical concepts and their respective mathematical representations, so that it is always clear what the mathematical structure represents physically and no mathematical objects are unphysical.

Another important difference is that we typically start at a lower level. For example, we do not assume probability spaces, which are already constructions upon sigma-algebras and measures, and sigma-algebras are typically constructed from the topology of the space. We instead aim to recover those as well. This is particularly important if one is interested in the structure of space-time, since its topology and sigma-algebra are the most foundational aspects.

In previous work\cite{Carc2}, we have shown the link with this idea to the mathematical framework of topology. That is, in physics, the topology captures what is experimentally verifiable (i.e. $U$ is an open set if and only if ``$x$ is in $U$'' is experimentally verifiable). For real valued quantities, the verifiable statements correspond to open intervals to signify that our measurements always have finite precision. The fact that a continuous quantity has precisely a given real value (e.g.~the length of the side is exactly 1 meter and the diagonal is exactly $\sqrt{2}$ meters) is not experimentally verifiable. This gives new insight to the notion of topological continuity and it tells why, in very general terms, functions \textbf{must} be well behaved in physics. It is not a matter of convenience: if they are not they would break experimental verifiability.

This idea also clarifies the division between topology and geometry. Topological constructs are more primitive than geometrical ones because they have no notion of size. Physically, it means that before we can assign distances between two points, we must be able to distinguish them, to tell them apart. The topology, then, comes first, as it tells us if and how the points are distinguished experimentally. This insight also tells us why different spaces, though topologically equivalent, have different geometry. In everyday space, the distance along any direction can be measured in the same unit, say meters. The geometry is Riemannian. On phase space, within a degree of freedom, units of position cannot be compared to units of momentum, therefore we cannot define angles. Yet, phase-space areas for a degree of freedom are proportional to the number of possible configurations. The fact that we can quantify areas for each degree of freedom is what gives us symplectic geometry. If we imagine for example the space of all possible blood work results, however, we have neither notion, and therefore we simply have a manifold that is not geometric.

For those interested in the ultimate structure of space-time these insights naturally lead to the following line of inquiry. Before we want to understand whether space-time is geometrically truly a Riemannian manifold, we need to understand whether space-time is topologically truly a manifold. A manifold, simply put, is a set of possible cases $X$ that can be experimentally identified by a set of real values. So, the question is: when can a set of possible cases, or possibilities for short, be experimentally identified by a real value? To be clear, the question is not when the results of our measurements are real numbers. The answer to this is simple: never. Our measurements are always finite precision. The question is when the cases we want to be able to distinguish are parameterized by the real line.

We will concentrate on a single real line (i.e.~total ordering) and not address higher dimensional spaces specifically. Since space-time in relativity is four dimensional, it would seem we are ignoring the most interesting cases, but it is not so. The issue is that time itself plays a special role: in every local frame, we must be able to use the time coordinate as the affine parameter for the evolution of a particle by writing $x^i=x^i(x^0)$. As we mentioned before, functions need to be, at the very least, topologically continuous or they would break the notion of experimental verifiability (i.e.~verifiable statements would not be mapped to other verifiable statements). This means that in any theory that includes time evolution, the topology one gives to time will severely constrain the topology of the space within which time evolution takes place. Moreover, in any theory of space-time we have the additional constraint that time and spatial coordinates can be mixed. Therefore the argument will work in reverse as well: the topology one gives to space has to be the same as the one of time because one may use spatial distance as a time parameter. For example, two spaceships drifting at constant velocity may use their spatial separation as a clock.

Even though this work is mathematically very technical and abstract, we strive to have a well understood map to more tangible and physical concepts; that is one of the main goals in our work. Therefore let us first give a summary of the results in physics terms, leaving the details to subsequent sections.

\subsection{Summary of results}

Mathematics offers many constructions of the real numbers from more primitive notions. Common techniques include completion through Cauchy sequences, using Dedekind cuts or adding constraints over an algebraic field. In fact, Faltin et al. state\cite{faltin1975real}:
\begin{quote}
Few mathematical structures have undergone as many revisions or have been presented in as many guises as the real numbers. Every generation re-examines the reals in the light of its values and mathematical objectives.
\end{quote}
Our present purpose is one such re-examination in the light of how these quantities are defined through experiments.

If we think about how quantities are measured, the general idea is that we are able to define references with preset amounts, and then compare our object to several references to find the ones that bound the value from above and below. For example, a ruler is a series of marks and measuring the position with it means finding the closest one. A clock is a series of ticks and measuring time means noting the tick right before the event. A balance scale compares the weight of an object to a few known ones. Intuitively, we can understand that a quantity is continuous if, in line of principle, we can prepare references ever closer to each other. Our aim is to make this notion precise, so we have a practical/operational definition of what we mean experimentally by a continuous quantity.

A single reference is something that allows us to distinguish a before and an after, a smaller and a greater, a higher and a lower. We can formalize a reference with our more basic notion of a verifiable statement: a reference gives us two verifiable statements, ``the object is before this reference'' and ``the object is after this reference''. This allows us to recast ordering relationship in terms of logical relationship. For example, if reference A is after reference B, then if ``the object is before reference B'' is true, ``the object is before reference A'' must also be true. And these logical relationships remain conceptually the same regardless of whether ``before'' means ``before in time'' or ``before along a spatial direction'', etc. These are the sort of basic ideas and language we use within our framework, and from these we work out a set of necessary and sufficient conditions that a set of references needs to satisfy so that they define a set of possible distinguishable cases, or possibilities for short, that is ordered like the real line.

The idea is that the quantities themselves are not a priori objects, but a construction built upon a set of references that form our system of measurement. A reference frame in space-time, then, is also a construction consisting of fixed elements (e.g. the stars, the borders of the experiment table, the elements of a timing system in a particle accelerator, ...) and signals exchanged among them. The topology of space-time is then an idealized characterization of the set of all possible such constructions, that is, all possible reference systems. Our goal is to understand the extent and the limitation of such idealization.

The biggest obstacle in this undertaking, and possibly in fully appreciating it, is that our standard intuition traps us in circular arguments. For example, one issue is that references have an extent: they occupy some space and so do objects. If we are to measure the position of an object, then, it may extend before, on and after the mark of a ruler. In these cases we typically note the position of the beginning of the object and of the end of the object. This implicitly relies on the fact that the resolution of our eyes is higher than that of the ruler, and that we can independently recognize parts of the object, namely the beginning and end. Identifying parts using a higher resolution corresponds, in the end, to new references. So, assuming this can always happen means implicitly assuming what we are tasked to derive.

Unfortunately, thinking visually and drawing pictures is misleading because everything we draw is inherently ordered. It confuses more than it helps, so much so that we ourselves progressed only when we stopped trying to reason with pictures and concentrated on just the boolean logic of our statements in truth tables. In that setting, the only elements we are allowed to define are those simple logical relationships we described before: if the object is before A, then it is not after B and it is before C. All properties and relationships between references, then, must be defined in that manner.

Using that strategy, we reached two major findings. The first major finding is that \emph{most of the conditions are not required by the real numbers specifically, but they are required by ordered quantities in general}. That is, the set of references has to satisfy a few conditions just to be sure that the cases they distinguish experimentally are ordered (i.e.~can be thought to be all being one after the other). The conditions are the following.
\begin{itemize}
	\item The references must be strict, meaning that the cases before, on and after are mutually exclusive. In practice, if objects have an extent, this must be much smaller than the extent of the reference, such that it will always be considered wholly before, on or after the reference.
	\item The references must be aligned, meaning their before and after statements identify incremental regions. In practice, this means we can always distinguish between them and they remain with constant before/after relationships (e.g.~they do not fluctuate). 
	\item The references must be refinable, meaning we can always resolve overlaps and fill in the whole space. For example, if an object can be between two references, then we must be able to put a new reference between those two references as well. If we have two references that overlap, then we must be able to find other references that fit within that overlap, to tell what is before and after.
\end{itemize}
These are the conditions one must require to have any linear order, either discrete (which can be labeled by integers), continuous (which can be labeled by real numbers) or any other. The divergence between real and integer quantities lies in how many references one can put in between two references. In the integer case, we can only fit finitely many references. In the real case, we can fit infinitely many.

The second finding is that \emph{the inability to distinguish below a certain scale most likely invalidates those conditions, leading to a breakdown in ordering.} In other words, it is not just that the points of our space cannot be labeled by real numbers: they cannot be ordered and therefore cannot be labeled by any ordered quantity. To obtain that result, one must simply argue that at least one of the three conditions fails. For example, one could argue that if at the finest level the references are particles, once they start being very close to each other it may become impossible to keep them distinguishable, which makes it impossible to keep well established before/after relationships and therefore satisfy the alignment condition. One could argue that if both references and the object being measured are particles, their extent (i.e.~the support of the wavefunction or of the field excitation) is comparable, which breaks the strictness condition. Alternatively, one can also argue that if we approach Planck scale, we cannot refine our references any longer, which breaks the refinement condition. The gist is that, in the case of quantities over which objects have an extent, the requirements for ordering can only be understood in terms of simplifying assumptions.

The inability to create an ordering at the finest scale means we are not able to define other geometrical quantities, such as distances and angles. In this light, the ultimate structure of space-time may not be geometrical but it will still be topological, as we will still need to describe what is experimentally accessible. At large scale, geometrical structure will need to emerge to recover the established theories. The point is that these large scale geometric structures cannot emerge from other geometrical structures, as these would necessarily suffer from the same fundamental problems.

\subsection{Outline}

The full mathematical account, which is available at \cite{Carc3}, uses tools and constructions that, in our experience, are not widespread among physicists. Moreover, that level of detail may not be of interest to most. Therefore, we have recast the definitions and the main arguments to the most basic elements of set theory and order theory, which we believe are more accessible and cover the most important points. This should allow us to give a more intelligible account of the physics without hiding it in the math.

We will first review the link between point-set topology and experimental verifiability. We will show how a reference can be defined in terms of sets. We will give the basic notions of order theory we need. We will then formalize the requirements a set of references needs to satisfy to identify a continuous quantity. Finally, we will see how these requirements describe idealized conditions.

\section{Elements of topology and its link to experimental verifiability}

The first thing we need to establish is the link between topology and experimental verifiability. Point-set topology (or general topology) is, knowingly or unknowingly, widely used in physics as it is the foundation of many other tools, such as differential geometry and Lie groups. Most mathematical structures used in physics are topological spaces. Therefore it is crucial to understand what physical content they capture.\footnote{In computer science, the link between topology and computability is already accepted, see e.g. \cite{escardo2004synthetic}, though not widely known. There is a link between those concepts and the ones presented here, but, for brevity, we are not going to expand on it. The gist is that any computation device is also a physical system, and the output of a computation can be experimentally verified (i.e.~we can read it).}

The formal definition of a topology is the following: given a set $X$, a topology $\mathsf{T}$ is a collection of subsets of $X$ that is closed under finite intersection and arbitrary (infinite) union.\footnote{Technically, it also must contain the full set $X$ and the empty set $\emptyset$. This does not play an important role.} This is similar to other algebraic structures, like groups, where you have elements and operations that return other elements of the same structure. Here the elements are subsets of $X$ and the operations are set operations. The formal definition is very abstract and does not have an apparent connection to physics.

Seemingly unrelated, consider a ``verifiable statement'': an assertion that can be verified to be true experimentally. For example, ``the mass of the neutrino is $0.05 \pm 0.005$ eV" or ``the electron is a negatively charged particle''. In general these statements do not perfectly identify a single possible case, rather they identify a set of possible cases. If $X$ is the set of all possible cases, each verifiable statement is associated with a set $U \subseteq X$ and the statement can be re-expressed as ``$x$ is in $U$''. In our first example, $x$ is the mass of the neutrino and $U$ is the set of possible values between $0.045$ and $0.055$.

Now we note that not all subsets of $X$ correspond to a verifiable statement. For example, ``the mass of the neutrino is exactly $0.05$ eV'' is not verifiable because we cannot perform a measurement with infinite precision. We call $U \subseteq X$ a verifiable set if the statement ``$x$ is in $U$'' is, at least in line of principle, something we can verify experimentally. Let $\mathsf{T}$ be the collection of all verifiable sets. The idea is that $\mathsf{T}$ is a topology.

To convince ourselves of that, consider the following. If we have a finite collection of verifiable sets $\{U_i\}_{i=1}^n$, then the intersection corresponds to the logical AND (i.e. the conjunction) of all the statements ``$x$ is in $U_i$''. We can test the intersection simply by testing each assertion one at a time, and therefore $\{U_i\}_{i=1}^n$ is a verifiable statement. If the collection were infinite, though, we would have to go through infinitely many tests to verify the logical AND, so we would never terminate. Therefore, in general, the infinite intersection of verifiable sets is not verifiable.

On the other hand, the union of verifiable sets would correspond to the logical OR (i.e. disjunction) of all the associated statements. In this case, as long as one statement is verified, the OR is verified and we can terminate. Therefore we do not care how many statements there are after the one that was verified, so the infinite union of verifiable sets is verifiable. Note that topologies are closed under arbitrary unions, not just countable. Yet, because we cannot test more than countably many statements, the set of all verifiable statements must be constructable for a countable set of statements, or we would not be able to fully explore the space even with unlimited time. Therefore the topology $\mathsf{T}$ must have a countable basis, which means all arbitrary unions are countable unions. 

We can make this more concrete by looking at the standard topology on the real numbers $\mathbb{R}$, which contains all the open intervals $(a,b)$ where $a,b \in \mathbb{R}\cup\{-\infty, +\infty\}$. These, in fact, correspond to all the finite precision measurements. It also contains their unions. For example ``the absolute value of the charge of the electron is $1.6 \pm 0.005\;10^{-19}$ C'' would correspond to ``$x$ is in $(-1.65 \;10^{-19}, -1.55 \;10^{-19})\cup(1.55 \;10^{-19},1.65 \;10^{-19})$". Singletons, sets with a single value, are not part of the topology and in fact we cannot verify them experimentally. Note that the topology can be generated by the set of all rational intervals, which is countable, since the infinite unions will allow us to construct limits, which correspond to the intervals of the reals.

A topology, then, is not just some mathematically abstract construction that happens to be useful in physics. It captures the way that we can experimentally distinguish a set of possible cases. The main point is that statements have a binary logic in terms of TRUE/FALSE, while experimental tests have a ternary logic in terms of SUCCESS/FAILURE/UNDEFINED, where undefined corresponds to non-termination. The topology is generally used to keep track of those differences. It makes sense, then, that topologies are so pervasive among the mathematical structures used in physics: experimental verifiability is at the heart of science.

Note that we will call verifiable sets, instead of open sets, the sets within the topology. We will also call falsifiable sets, instead of closed sets, the complement of verifiable sets. It will make our work more connected to this physical notion and therefore a little bit more intuitive, especially to those who are not deeply familiar with point-set topology.

The question we want to answer is the following: how are the real numbers constructed in physics? What are the requirements on our experimental apparatus such that those quantities can be measured? Can we expect those requirements to hold when we go to ever smaller scales? If not, how do they break down?

Since we established a link between topology and experimental verifiability, we can pose the above questions in a way that is both mathematically and physically precise. Suppose we have a set of possible cases $X$, which we will call \textbf{possibilities}, and a way to distinguish them experimentally, which means we have a topology $\mathsf{T}$ that captures all our verifiable sets. What are the requirements on $(X, \mathsf{T})$ such that it is homeomorphic (i.e. topologically equivalent) to the set of the real numbers with standard topology?

\section{References as the starting point for measurement scales}

The first thing we need to develop is a conceptual model to represent how quantities are measured experimentally. Let us first gather some requirements from the physics.

We start from the notion of a reference, such as a mark on a ruler or the tick of a clock. A reference, then, is a physical object that allows us to distinguish between three cases: a before, an on and an after the reference. That is, a point can be before, on or after the mark on a ruler; an event can happen before, on or after the tick of a clock. We take references to be the basic conceptual element upon which quantitative measuring devices are constructed in practice.

In general, the three cases a reference defines are not mutually exclusive: an object can extend before, on and after the mark; an event can start before and end after the tick of a clock. In fact, if the object is both before and after, it will be on the reference as well. In all cases, the object should be found at least in one of the three cases.

If a reference is at an end of the range, then it will either not have an after or a before. Yet, a reference is a physical object and needs to take some space, therefore there will always be an on case.

We should also note that the before and after cases are easier to test experimentally. When comparing two weights on a scale, for example, it is easy to tell when one is greater than the other. If they are very close, we can only typically say that they are closer than a certain threshold. This is also something we will have to take into account.

If $(X, \mathsf{T})$ is the topological space of the physically distinguishable cases, the number of ways an object can be found to be, we can formalize a reference as a triplet $r = (B, O, A)$ of three subsets $B, O, A \subseteq X$. Respectively, they will represent the cases in which the object is before, on or after the reference. We can capture the rest of the requirements with the following:
\begin{itemize}
	\item $B \cap A \subseteq O$ (all possibilities in which the object is found before and after, the object is also found on)
	\item $B \cup O \cup A = X$ (before/on/after cover all possibilities)
	\item $O \neq \emptyset$ (the reference itself must be found somewhere)
	\item $B, A \in \mathsf{T}$ (before and after are verifiable)
\end{itemize}
Note that we require only before and after to be experimentally verifiable, leaving the on case unspecified. It turns out this is sufficient and avoids the problem of testing for perfect equality.

A measuring device will be composed of several references. Intuitively, for continuous quantities we are assuming that, though we always have a finite set of references with finite precision, we could in principle refine our instruments with ever greater precision. Our work is to make this intuitive notion precise.

\section{Elements of order theory}

There are many ways to mathematically characterize the real numbers. The one we are interested in is in terms of their order, instead of operations like addition or multiplication, as it is the one that imposes the least requirements and it is enough to identify the topology.\footnote{Note how all transformations that preserve addition or multiplication must preserve the ordering, while the converse is not true. For example, consider $x \to x^3$, which is a non-linear monotonic transformation. It preserves the ordering (i.e. $x_1 \leq x_2$ if and only if $x_1^3 \leq x_2^3$) while it does not preserve addition (i.e. $(x_1 + x_2)^3 \neq x_1^3 + x_2^3$).} Let us review, then, a few key concepts of order theory, which is the branch of mathematics that studies ordered sets and their properties.

A partial order $\leq$ is a binary relationship that is reflexive (i.e. $a \leq a$), antisymmetric (i.e. if $a\leq b$ and $b \leq a$ then $a = b$) and transitive (i.e. if $a \leq b$ and $b \leq c$ then $a \leq c$). In a partial order, two distinct elements are not necessarily one before the other. On a plane, we can order points by their horizontal position, yet that would not order points that lie on a vertical line. A total order, or linear order, is a partial order such that any two elements are comparable (i.e. at least $a \leq b$ or $b \leq a$). If not explicitly stated, we will assume orders to be linear.

Given an ordered set $(X, \leq)$ we can construct the order topology in the following way. Take all sets of the form $(a, \infty) = \{x \in X \, | \, x > a\} \;,\; (-\infty, b) = \{x \in X \, | \, x < b\}$. This will be a basis for the order topology. We then take all the sets that can be generated from the basis through finite intersection and arbitrary union. This will be the order topology. Both the integers and the reals are totally ordered sets with their standard $\leq$ relationship. Their respective order topologies correspond to their standard topologies. Note how each set within the basis corresponds to a statement like ``$x$ is after $a$'' or ``$x$ is before $b$''. Note how these, already, are very similar to the before and after cases we introduced in the previous section.

Like groups or topological spaces, two ordered sets are isomorphic if they have equivalent structure. That is, two ordered sets are isomorphic if there is a bijection that preserves the ordering (i.e.~an invertible monotonic function). If two ordered sets are isomorphic, then they will have the same order topology and vice-versa. So the ordering identifies the topology of an ordered set and vice-versa.

In light of this, we should break our problem into two parts. First understand what the requirements are on references such that the set of possibilities $X$ is linearly ordered and its topology is the order topology. Then understand what the additional requirements are to be ordered like the real numbers or the integers. It turns out that most of the work lies in the first part, in recovering the linear order.

\section{Experimental requirements for constructing real-valued quantities}

\subsection{Strict references}

As we saw, in general a reference defines three cases (before/on/after) that may not be mutually exclusive. We say a reference is strict if they are. That is, a reference $r= (B, O, A)$ is strict if $B \cap O = \emptyset$, $O \cap A = \emptyset$ and $B \cap A = \emptyset$. Technically, we would just need that before and after are disjoint ($B \cap A = \emptyset$) and then redefine on as what remains ($O = X \setminus (A \cup B)$). That is, as long as before and after are separate cases, we can redefine on as ``not before and not after''.

The order topology always allows us to construct strict references. If we take $a,b \in \mathbb{R}$ such that $b<a$, we have $( (-\infty, b), [b, a], (a, +\infty) )$ that represents a reference that physically extends from $b$ to $a$. Mathematically, we can see that the extent of the reference is represented by a falsifiable (i.e. closed) set $[b,a]$, while the others are verifiable (i.e. open) sets. Therefore if we want to rederive the order topology starting from references alone, those will need to be strict. Intuitively, defining an order experimentally on all possibilities means that, given two distinct ones, we are able to find references that will confirm unambiguously that one possibility is before the other. Therefore we must be able to tell before and after apart.

What does this requirement entail in practice? References for quantities that we count (i.e.~discrete quantities) can always be made strict: either you have $n$ elements, less than $n$ or more than $n$. The reason is that the quantity does not really have an extent and the possible values are well separated. Quantities over which we have an extent, like space and time, are another matter. When measuring the position of an object, this can extend before, on and after the mark. If the extent of the object being measured is much smaller than the extent of the reference, then we can assume the object to be wholly found either before, on or after. In those regimes we can treat the references as strict. But this is an idealization that will break down if the extent of the references is comparable to the extent of what is being measured.

\subsection{Aligned references}

As we put references together, we must be sure they are related to each other in the correct way if we want to end up with a linear ordering. Intuitively, if we mix references for horizontal position with references for vertical position we will not end up with a linear ordering.

To define what it means for two references to be aligned, take two values $b_1, b_2 \in \mathbb{R}$. Suppose that $b_1 \leq b_2$. Then if we find that our object is before $b_1$ then it will also be before $b_2$. In terms of sets, $(-\infty, b_1) \subseteq (-\infty, b_2)$. In a linear order, the idea that one point is before the other can be translated in terms of set inclusion. Alignment between references, then, can be defined by requiring that the before and after statements have an inclusion relationship.

We also note that the negation of ``$x$ is greater than $a$'' is ``$x$ is less than or equal to $a$''. In terms of sets, $(a, +\infty)^C = (-\infty, a]$. For a strict reference extending from $b$ to $a$, we will have $(-\infty, b) \subseteq (-\infty, a]$.

We will say that two references are aligned if their edges can be ordered. Mathematically two references $r_1 = (B_1, O_1, A_1)$ and $r_2 = (B_2, O_2, A_2)$ are aligned if the sets $B_1, B_2, A_1^C, A_2^C$ can be ordered by inclusion. For example, if $B_1 \subset A_1^C \subset B_2 \subset A_2^C$ then we will be in the case where $r_1$ is completely before $r_2$. If $B_1 \subset B_2 \subset A_1^C \subset A_2^C$ then the two references are overlapping. Alignment of references is another necessary condition if we want to reconstruct an ordering that is linear.

What does this requirement entail in practice? It means that there are clear fixed ordering relationships between the references. For example, every time something is before one reference it will also be after the other. It means we are perfectly able to prepare, control and identify our references.

To give a better understanding of how these definitions work in a multidimensional setting, let us see how they apply to a Cartesian frame with coordinates $(x, y, z)$. A reference for the $x$ coordinate would be something like $r_1=$(``$x < 0$'',``$0 \leq x \leq 2$'', ``$ x > 2$''). Note that it partitions the whole space in three disjoint regions. The on case acts as a divider and extends throughout $y$ and $z$. The reference $r_2=$(``$y < 0$'',``$0 \leq y \leq 1$'', ``$y > 1$'') is not aligned with $r_1$ as something can be before $r_1$ and still be before, on or after $r_2$. On the other hand, $r_3=$(``$x < -1$'',``$-1 \leq x \leq 1$'', ``$x > 1$'') is aligned with $r_1$ even though it is not either before or after $r_1$ because the on regions overlap. Note that these references would not have such a simple expression in, say, spherical coordinates. Yet the regions themselves are coordinate independent and, therefore, the relationships between them are too.

What happens is that each reference frame would have its own set of references that can be nicely expressed using its coordinates. The before/after relationships expressed using one set of references may not be simply expressible using another, precisely because they are not aligned with each other. Relativistically, the references used for time by one frame will not be aligned with the ones of a boosted frame: the coordinate surfaces do not partition (i.e. do not foliate) space-time in the same way. Reference alignment essentially captures these various requirements with one simple formal definition.

\subsection{Refinable references}

Strict and aligned references allow us to define ordering between sets of possibilities. We need an additional condition to make sure we have enough references at an appropriate resolution to be able to order the possibilities themselves. That is, to order all singletons, the sets with only one possibility.

The general idea is we must be able to fill in the gaps and break apart overlapping regions. For example, if we have two references that can have something in between, then we must be able to put a reference between them. That is, if $r_1 = (B_1, O_1, A_1)$ and $r_2 = (B_2, O_2, A_2)$ are such that $A_1 \cap B_2 \neq \emptyset$, then we can find a reference $r_3 = (B_3, O_3, A_3)$ such that $O_3 \subseteq A_1 \cap B_2$. On the other hand, if the extent of one reference is within the extent of the other, then we need to be able to find a reference that covers another part. That is, if $O_2 \subset O_1$ then we can find another reference such that $O_3 \subset O_1$ and $O_2 \cap O_3 = \emptyset$. If a set of references has this property, then we say it is refinable: the references can be refined to non-overlapping references that cover all the possibilities.

What does this requirement entail in practice? It means that we can place a reference wherever we want and can shrink its extent such that it occupies only one possibility. If we take the traditional manifold structure of space-time literally, this would mean having at our disposal ever shrinking references that can be placed anywhere in space and time. For example, it would require being able to create a timing system with as many well separated pulses as desired distributed to as many places as desired. All of this, without changing the nature of the process we are studying.

\subsection{Linear order}

It can be demonstrated that a set of experimentally distinguishable possibilities is linearly ordered if and only if its topology can be generated by a set of refinable, aligned strict references. That is, if we have an order topology, we can construct a set of references that have those properties, and if we have a set of references that have those properties, the order topology can be generated by finite intersection and arbitrary union from the before and after sets of the references.

The idea is that, under those conditions, we can find references that are so fine that they extend over only one possibility. Therefore we have a one-to-one correspondence between experimentally distinguishable cases and the finest references. Since all references are aligned and strict, two different references must be sequential, one before the other. This order corresponds to the one that defines the order topology.

\subsection{Discrete (integer) order}

The order of the integers can be characterized as the one that does not have a minimum, does not have a maximum, and given any two elements, there are finitely many elements between them.

If we have an ordered set with the same characteristics, we can put it in a bijective correspondence with the integers that preserves the order. Roughly, one can proceed this way. Take one element and arbitrarily label it zero. Since between two elements there are only finitely many, each element will have a successor and a predecessor. Label one the successor of zero and label minus one the predecessor of zero. Then label two the successor of one, and so on. Since between zero and any other element there are only finitely many other elements, at some point we will reach all the elements.

If we have a set of refinable aligned strict references, then the only additional requirement needed to recover the integers is that between two references there are only finitely many references. The finest possible references, which will correspond to the possibilities of the space, can then be labeled by the integers. The possibilities are ordered like the integers.

\subsection{Continuous (real) order}

The order of the reals can be characterized as one that is dense (i.e.~between two elements there is always another one), complete (i.e.~it contains all the limit points), has a countable dense subset (i.e.~between two real numbers we can always find a rational, which is a countable set), and has no minimum or maximum. 

If we have an ordered set $X$ with the same characteristics, we can put it in a bijective correspondence with the reals that preserves the order. Roughly, one can proceed this way. We start by noticing that we have a countable dense subset $Q$ and, as Cantor showed, it can be put into correspondence with the rationals. The set $Q$ is dense in the original set $X$ and the original set $X$ is complete. Therefore $X$ is the completion of $Q$. Another theorem in order theory states that the completion of any order is unique. Since the real numbers are the completion of the rationals, and $Q$ has the same order of the rationals, then $X$ has the same order of the reals.

If we have a set of refinable aligned strict references, then the only additional requirement needed to recover the reals is that between two references there is always another reference. That is, we are only imposing that the references are dense. The completion and the countable dense subset are a consequence of the more general topological requirements. The inclusion of all the limit points will come from the infinite union allowed by the topology while the dense subset is associated with the countable base, which is a requirement for our topology to be physically meaningful.



\section{Breakdown of ordering and of geometry}

In the previous sections we have identified the experimental requirements needed to be able to operationally define continuous quantities. These are:
\begin{enumerate}
	\item the ability to compare the quantity (i.e.~position, time, mass, ...) with references such that we can say whether the object is before or after each one
	\item the three different cases, before/on/after, must be mutually exclusive
	\item the references must be aligned, we must be able to place them in a line
	\item we must be able to find enough references such that overlaps can be avoided and all the possible values can be covered
	\item between two references we can always place another
\end{enumerate}
Under these conditions, we can assign a real number to the finest possible references, which will correspond to all possible distinguishable cases. Higher dimensions can be constructed by combining sets of references for different quantities, which leads to the familiar structure of a manifold.

We want to stress that these requirements are not properties of the quantity being measured. They are operational requirements. They correspond to the mundane act of constructing a ruler, a calorimeter, a timing system, or any other experimental device.

We are not, therefore, positing that there already exists a continuous quantity to be found. We are asking under what conditions we can label the outcomes of our measurement scheme with a continuous quantity. This is in line with the perspective of considering a physical theory, including quantum theory, as dealing with the output of measurements. For example, Wheeler commented\cite{Wheeler}: ``It is wrong, moreover, to regard this or that physical quantity as sitting out there with this or that numerical value''. What we are doing is going to a more fundamental level to understand how this works in the continuous case, and how can it be modeled in a rigorous say.

Now that we understand what the requirements are we can ask: are these requirements tenable? Or are they necessarily idealizations?

Note that, with the exception of (v), all other requirements are needed for ordering itself. That is, the failure of any one of the first four means that ordered quantities are not physically meaningful. This means we cannot substitute the real numbers with the rationals or the integers, as these are themselves ordered. Note that all geometrical structures are based on some notion of numeric distance, therefore if order fails geometrical structures fail as well. We are not going to be able to define metric tensors, symplectic forms or even a differentiable structure.

As such, the bar to show that, at a fine level, space-time does not possess a geometrical structure is very low. One simply has to instill the doubt that any of (i-iv) is untenable. Here we give a few angles one may take. We will concentrate on measurements of spatial position. Similar arguments can be made for time as well.

For (ii), we saw that references are objects that have, in general, an extent in space. A reference can be considered strict if the extent of the object measured is much smaller than the extent of the reference. Ultimately, however, the finest constituents will also be the finest objects that we can use as references. Therefore, if we assume that we can make finer and finer references, at some point we will find the extent of the references comparable to the extent of the objects we are measuring. This is also true for fundamental particles, where the extent of the system corresponds to the extent of the wave function: if the wavefunctions overlap, one particle is not clearly wholly before/on/after the other. Similar arguments apply for excitations of quantum fields.

For (iii), the requirement for two references to be aligned means, for example, that we must always be sure which one is before the other. This means we must have a way to tell the references apart. This is problematic if the fundamental objects are ultimately indistinguishable. We cannot order what we cannot distinguish.

For (iv), the requirement for refinement means we can always prepare finer and finer references. This is generally thought not to be possible once length scales are comparable to the Planck length.

We are not interested here to expand on these arguments or pick ones that are universally accepted. That will be the scope of future work. Here we simply want to show that it is relatively easy to formulate reasonable arguments. As long as one finds reasonable that one of (i-iv) fails, then our point follows: space-time structure at a fine scale will not be geometrical, will not be ordered, but it will still be topological, giving us some way to experimentally distinguish what can physically be well defined.

\section{Conclusion}

We have seen what the requirements are to give rise to the real numbers through a set of experimentally verifiable statements, which captures the most basic elements of how continuous quantities are measured in practice. We have also seen that those can only be considered idealized conditions, and that, when those requirements fail, ordering itself fails and no geometrical structure can be constructed. Naturally, this opens the question of what topological spaces would be appropriate in those regimes. This is something we do not have an answer for, though we can mention two of the possibly many scenarios.

One can imagine to try and construct a topological space that is not ordered at a fine scale but is ordered at a large scale. A difficulty here is that typically the notion of distance is absent from topological spaces. Though it is not clear to us whether this is possible, at least it is a clear venue to explore.

There is, however, a much more drastic scenario. General relativity tells us that the geometry depends on the content of the space (i.e.~the energy-momentum tensor). What is in the space affects the notion of distance and angle. It would not be far-fetched to assume that what is in the space also affects what can be distinguished experimentally. As the energy-matter distribution changes the geometry by curving space-time, it would, at a more fundamental level, change the topology as well. It is not clear to us at this time how one would even start exploring such a scenario.

\section*{Acknowledgments}

We would like to thank Mark J. Greenfield for review and help on the mathematical details. Funding for this work was provided in part by the MCubed program of the University of Michigan. This article is part of a larger project, Assumptions of Physics, that aims to identify a handful of physical principles from which the basic laws can be rigorously derived~\cite{Carc3}.

\section*{References}

\bibliographystyle{unsrt}
\bibliography{bibliography}

\begin{thebibliography}{10}

\bibitem{PhysRevA.84.012311}
Giulio Chiribella, Giacomo~Mauro D'Ariano, and Paolo Perinotti.
\newblock Informational derivation of quantum theory.
\newblock {\em Phys. Rev. A}, 84:012311, Jul 2011.

\bibitem{QLogicReview}
Bob Coecke, David Moore, and Alexander Wilce.
\newblock {\em Current Research in Operational Quantum Logic: Algebras,
  Categories, Languages}.
\newblock Springer, Dordrecht, 01 2000.

\bibitem{Hardy:2001jk}
Lucien Hardy.
\newblock {Quantum theory from five reasonable axioms}.
\newblock 2001.
\newblock quant-ph/0101012.

\bibitem{ludwig_hein_2013}
G.~Ludwig and C.~A. Hein.
\newblock {\em Foundations of Quantum Mechanics}.
\newblock Springer Berlin, 2013.

\bibitem{Carc1}
Gabriele Carcassi, Christine~A. Aidala, David~J. Baker, and Lydia Bieri.
\newblock From physical assumptions to classical and quantum {Hamiltonian} and
  {Lagrangian} particle mechanics.
\newblock {\em Journal of Physics Communications}, 2(4):045026, apr 2018.

\bibitem{Carc2}
Christine~A Aidala, Gabriele Carcassi, and Mark~J Greenfield.
\newblock Topology and experimental distinguishability.
\newblock {\em Topology Proceedings}, 54:271--282, 2019.

\bibitem{faltin1975real}
F~Faltin, N~Metropolis, B~Ross, and G-C Rota.
\newblock The real numbers as a wreath product.
\newblock {\em Advances in Mathematics}, 16(3):278--304, 1975.

\bibitem{Carc3}
G.~Carcassi and C.~A. Aidala.
\newblock Assumptions of physics.
\newblock In preparation, available from http://assumptionsofphysics.org/book,
  2019.

\bibitem{escardo2004synthetic}
Mart{\'\i}n Escard{\'o}.
\newblock Synthetic topology: of data types and classical spaces.
\newblock {\em Electronic Notes in Theoretical Computer Science}, 87:21--156,
  2004.

\bibitem{Wheeler}
John~A. Wheeler.
\newblock Information, physics, quantum: The search for links.
\newblock {\em Proceedings of 3rd International Symposium on the Foundations of
  Quantum Mechanics}, pages 354--368, 1989.

\end{thebibliography}

\end{document}